\documentclass[aps,twocolumn,showpacs,showkeys]{revtex4}
\usepackage{graphicx}
\usepackage{amssymb}
\usepackage{bm}
\begin{document}
\setcounter{page}{1}
\title[]{Effects of selective dilution on the magnetic properties\\ of
 La$_{0.7}$Sr$_{0.3}$Mn$_{1-x}$\textit{M}$^\prime_x$O$_3$ (\textit{M}$^\prime$ = Al, Ti)}
\author{D. N. H. \surname{Nam}}
\email{daonhnam@yahoo.com}
\thanks{Fax: +84-4-836-4403}
\author{N. V. \surname{Dai}}
\author{L. V. \surname{Hong}}
\author{N. X. \surname{Phuc}}
\affiliation{Institute of Materials Science, VAST, 18 Hoang-Quoc-Viet, Hanoi, Vietnam}
\author{L. V. \surname{Bau}}
\affiliation{Department of Science and Technology, Hongduc University, Thanhhoa, Vietnam}
\author{P. \surname{Nordblad}}
\affiliation{The \AA ngstr\"{o}m Laboratory, Uppsala University, Box 534, SE 751-21 Uppsala, Sweden}
\author{R. S. \surname{Newrock}}
\affiliation{Department of Physics, University of Cincinnati, OH 45221-0011, USA}
\date[]{Received \today}

\begin{abstract}
The magnetic lattice of mixed-valence Mn ions in La$_{0.7}$Sr$_{0.3}$MnO$_{3}$ is selectively diluted by partial substitution of Al or Ti for Mn. The ferromagnetic transition temperature $T_\mathrm{c}$ and the saturation magnetization $M_\mathrm{s}$ both decrease with substitution. By presenting the data in terms of selective dilution, $T_\mathrm{c}$ in the low-doping region is found to follow the relation $T_\mathrm{c}=T_\mathrm{c0}(1-n_\mathrm{p})$, where $T_\mathrm{c0}$ refers to the undiluted system and $n_\mathrm{p}$ is the dilution concentration defined as $n_\mathrm{p}=x/0.7$ or $n_\mathrm{p}=x/0.3$ for $M^\prime=$ Al or Ti, respectively. The scaling behavior of $T_\mathrm{c}(n_\mathrm{p})$ can be analyzed in the framework of the molecular-field theory and still valid when Mn is substituted by both Al and Ti. The results are discussed with respect to the contributions from ferromagnetic double exchange and other possible antiferromagnetic superexchange interactions coexisting in the material.
\end{abstract}



\maketitle

\section{INTRODUCTION}

Perovskite manganites with the general composition $(R,A)$MnO$_3$ ($R$: rare earth, $A$: alkali elements) have been intensively studied in the last decade since the discovery of the Colossal Magneto-Resistance (CMR) phenomenon in La$_{2/3}$Ba$_{1/3}$MnO$_3$ thin films \cite{Helmolt}. Manganite materials are interesting not only for fundamental researches, but also for practical applications. Along with sophisticated measurement and sample preparation techniques, chemical substitution has been widely used as a convenient method to uncover the underlying physics and to search for compositions with novel properties. Results for both $R$- and Mn-site substitution have been quite well documented in the literature. Magnetic and transport properties of manganites have been found to be systematically influenced by the average ionic size and the size mismatch of the ions occupying $R$-site \cite{Hwang,Damay,Martinez,Arulraj0}. However, substitution at Mn-site would lead to a mixture of effects caused by changes such as of crystal structure, charge carrier concentration, and more importantly the interaction between Mn and the substitutional ion. Despite many attempts at modifying the magnetic lattice of Mn ions by direct substitution at the Mn-site, no universal features have been revealed to date. In this paper, we report that the reduction of the ferromagnetic (FM) ordering transition temperature $T_\mathrm{c}$ of La$_{0.7}$Sr$_{0.3}$Mn$_{1-x}$\textit{M}$^\prime_x$O$_3$ (\textit{M}$^\prime$ = Al, Ti) scales with the relative substitution concentration and is consistent with a prediction from molecular-field theory (MFT). The substitution with Al or Ti is selective in nature because Al$^{3+}$ would only substitute for Mn$^{3+}$ and Ti$^{4+}$ for Mn$^{4+}$. Besides, because Al$^{3+}$ and Ti$^{4+}$ ions do not carry a magnetic moment, they are expected not to participate in any magnetic interaction.

\section{Experiments}
\begin{figure}[t!]
\includegraphics[width=3.in]{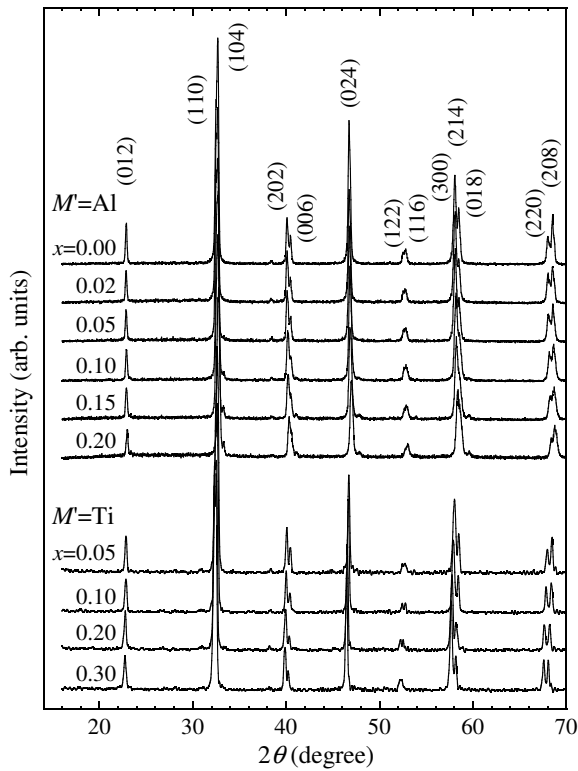}
\caption{Room-temperature x-ray diffraction patterns of La$_{0.7}$Sr$_{0.3}$Mn$_{1-x}M^\prime_x$O$_3$ ($M^\prime =$ Al: upper panel, $M^\prime=$ Ti: lower panel).} \label{fig.1}
\end{figure}

The La$_{0.7}$Sr$_{0.3}$Mn$_{1-x}$\textit{M}$^\prime_x$O$_3$ ($x=0-0.2$ for \textit{M}$^\prime =$ Al and $x=0-0.3$ for \textit{M}$^\prime =$ Ti) samples were prepared by a conventional solid state reaction method. Raw powders with appropriate amounts of high purity La$_2$O$_3$, SrCO$_3$, MnO$_2$, Al$_2$O$_3$, and TiO$_2$ are thoroughly ground, mixed, pelletized, and then calcined at several processing steps at temperatures increased from 900 $^\mathrm{o}$C to 1200 $^\mathrm{o}$C with intermediate grindings and pelletizations. The products are then sintered at 1370 $^\mathrm{o}$C for 48 h in ambient atmosphere followed by a very slow cooling process from the sintering to room temperature with an annealing step at 700 $^\mathrm{o}$C for a few hours. Room-temperature x-ray diffraction patterns (measured using a SIEMENS-D5000 with Cu-K$_{\alpha}$ radiation) presented in Fig. \ref{fig.1} show that all of the obtained samples are essentially single phase with perovskite rhombohedral (space group $R\overline{3}c$) structures, in agreement with previous studies \cite{Qin,Hu,Kallel,Kim,Troyanchuk}. Redox titration experiments (using K$_2$Cr$_2$O$_7$ titrant and C$_{24}$H$_{20}$BaN$_2$O$_6$S$_2$ as the colorimetric indicator) for checking the oxygen content, $3\pm\delta$, indicate that the samples are stoichiometric without any significant oxygen excess or deficiency. Magnetic measurements were performed in a Quantum Design MPMS-XL SQUID magnetometer.

\section{Results and discussion}

\begin{figure}[t!]
\includegraphics[width=3.in]{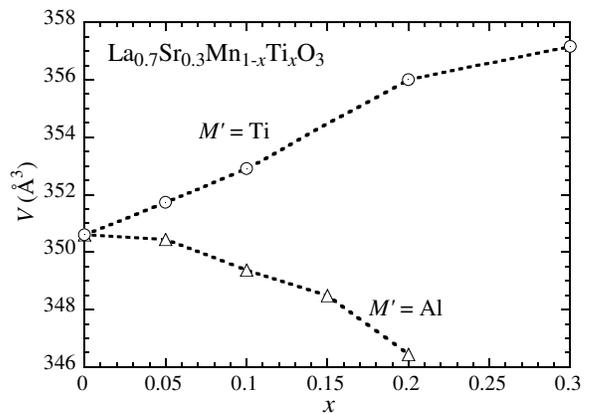}
\caption{Unit cell volume plotted as functions of Al and Ti concentrations. The data evidence a selective substitution of Al$^{3+}$ and Ti$^{4+}$ for Mn$^{3+}$ and Mn$^{4+}$, respectively.} \label{fig.2}
\end{figure}

There have been concerns in several previous studies \cite{Blasco,Krishnan,Nair} that Al$^{3+}$ could replace Mn$^{4+}$, causing oxygen deficiency in Al-doped manganites. However, our redox titration experiments carried out for the Al-substituted samples show that, except for the $x = 0$ and $x = 0.1$ samples that have slight oxygen excesses, the oxygen deficiency is negligibly small ($\delta\leq0.006$) and varies randomly with the substitution. Ti is generally believed to be tetravalent when doped in manganites. Both Ti$^{4+}$ and Mn$^{4+}$ are not Jahn-Teller active in the octahedral crystalline electric field. Liu \textit{et al.} \cite{LiuX} reported that Ti$^{4+}$ replaces Mn$^{4+}$ to form Ti$^{4+}$O$_6$ octahedra which are, as same as Mn$^{4+}$O$_6$ octahedra, not Jahn-Teller distorted, as evidenced in the infrared transmission spectra of the La$_{0.7}$Ca$_{0.3}$Mn$_{1-x}$Ti$_x$O$_3$ compounds. Alternatively, analyzes on the x-ray diffraction spectra of our La$_{0.7}$Sr$_{0.3}$Mn$_{1-x}$\textit{M}$^\prime_x$O$_3$ samples show that, as presented in Fig. \ref{fig.2}, the unit cell volume monotonically decreases with Al substitution while it increases with increasing Ti concentration. These structural evolutions clearly evidence a selective substitution of Al$^{3+}$ and Ti$^{4+}$ for Mn$^{3+}$ and Mn$^{4+}$, respectively, considering the fact that the ionic radius of Al$^{3+}$ (0.535 {\AA}) is smaller than that of Mn$^{3+}$ (0.645 \AA) and Ti$^{4+}$ (0.605 \AA) is larger than Mn$^{4+}$ (0.530 \AA) \cite{Shannon}.

As expected, the saturation magnetization, $M_\mathrm{s}$, measured at $T = 5$ K and in an applied field up to $H = 4$ T (not shown) decreases with increasing substitution concentration in both Al- and Ti-substitution series. In the high-field regime ($H\geq1$ T), the $M(H)$ curves are quite linear with a slope that develops with the substitution. For most cases, $M_\mathrm{s}$, determined either at $H = 4$ T or by extrapolating the high-field linear portion of the $M(H)$ curves to $H = 0$, does not reach and even further deviates from the theoretical value with increasing $x$. These observations suggest that, along with the major FM phase established by double exchange (DE) \cite{Zener}, there exists a non-FM phase that develops with the substitution at the expense of the FM phase. Since the Mn$^{3+}$/Mn$^{4+}$ ionic concentration ratio is driven off the optimal value of 7/3, a certain amount of the Mn ions may become redundant with respect to the DE couplings. That implies that the selective substitution on one Mn ionic species produces a redundancy on the other Mn species and those redundant ions contribute to the non-FM phase.

Temperature-dependent magnetization, $M(T)$, measurements show that the substitution of Al or Ti for Mn causes $T_\mathrm{c}$ to drop drastically (see the inset of Fig. \ref{fig.3} below). There have been several explanations for the reduction of $T_\mathrm{c}$ in other similar cases \cite{Qin,Hu,Kallel,Kim,Sawaki}. One explanation is simply that this is due to the suppression of long-range FM order of the localized $t_\mathrm{2g}$ spins by local breakdown of the exchange couplings where the substitution occurs \cite{Qin,Sawaki}. Hu \textit{et al.} \cite{Hu} assumed a demolition of the DE Mn$^{3+}$-O$^{2-}$-Mn$^{4+}$ bonds and a lower hole-carrier concentration caused by Ti substitution. Kallel \textit{et al.} \cite{Kallel} suggested that the presence of Ti favors the superexchange (SE) interaction and suppresses the DE mechanism. It has been also quite common to use the empirical formula of electron bandwidth \cite{Medarde},
\begin{equation}
W\approx\frac{cos[\frac{1}{2}(\pi-\theta_{\langle \mathrm{Mn-O-Mn}\rangle})]}{d^{3.5}_{\langle \mathrm{Mn-O}\rangle}},
\label{eqn1}
\end{equation}
 where $\theta_{\langle \mathrm{Mn-O-Mn}\rangle}$ and $d_{\langle \mathrm{Mn-O}\rangle}$ denote the average Mn-O-Mn bond angle and Mn-O bond length, respectively, to discuss the magnetic and transport properties of manganites \cite{Kim,Radaelli,Garcia,Arulraj,Das,Ulyanov}. Kim \textit{et al.} \cite{Kim} found that Ti substitution in La$_{0.7}$Sr$_{0.3}$Mn$_{1-x}$Ti$_x$O$_3$ increases the Mn-O-Mn bond length and reduces the bond angle. Based on structural data, the authors calculated the variation of $W$, finding a decrease of $W$ with $x$, and related it to the decrease of $T_\mathrm{c}$. Despite this consistence between $W$ and $T_\mathrm{c}$ found for the Ti-substituted samples, taking into account the ionic size differences and considering the opposite evolutions of crystalline structure presented in Fig. \ref{fig.2}, we proposed that the variation of $W$ may not be consistent with the decrease of $T_\mathrm{c}$ in the case when Mn is substituted by Al. This is verified in Fig. \ref{fig.4} where the variation of $W$ is presented for the La$_{0.7}$Sr$_{0.3}$Mn$_{1-x}$Al$_x$O$_3$ samples, showing a monotonic increase, in contrast to the decrease of $T_\mathrm{c}$ against Al concentration. We hereby emphasize that the variation of $W$ alone cannot explain the variation of $T_\mathrm{c}$ in $B$-site diluted manganites, even qualitatively.

\begin{figure}[t!]
\includegraphics[width=3.in]{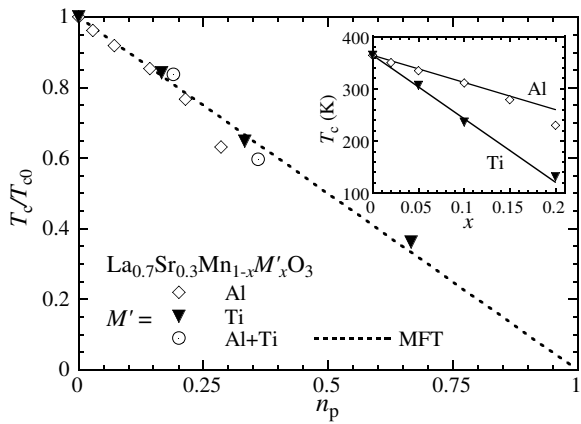}
\caption{The effect of selective dilution on $T_\mathrm{c}$ of La$_{0.7}$Sr$_{0.3}$Mn$_{1-x}M^\prime_x$O$_3$. $n_\mathrm{p} = x/0.7$ for $M^\prime=$ Al or $x/0.3$ for Ti. The $\odot$-symbols are for $M^\prime=$ Al$_{0.07}$Ti$_{0.03}$ and $M^\prime =$ Al$_{0.14}$Ti$_{0.06}$. The inset shows original $T_\mathrm{c}$ vs $x$ data.} \label{fig.3}
\end{figure}

The variations of $T_\mathrm{c}$ with $x$ for both series are presented in the inset of Fig. \ref{fig.3}. The effect of substitution with Ti on the reduction of $T_\mathrm{c}$ [defined by  $T_\mathrm{c} = T_\mathrm{c}(x)-T_\mathrm{c}(0)$] is as much as more than twice that of Al substitution. However, because of the selectivity of the substitution, it is more appropriate for $T_\mathrm{c}$ to be presented vs the relative substitution concentration $n_\mathrm{p} = x/0.3$ or $x/0.7$ for $M^{\prime} =$ Al or Ti, respectively. Interestingly, as can be seen in the main figure of Fig. \ref{fig.3}, the $T_\mathrm{c}(n_\mathrm{p})$ data collapse on the same linear curve that follows very well the relation $T_\mathrm{c} = T_\mathrm{c0}(1-n_\mathrm{p})$, being in consistent with a prediction derived from the molecular-field theory. According to MFT, in a system where a magnetic ion with spin $S$ interacts with its $z$ nearest-neighbor ions with an exchange coupling constant $J$, the ordering temperature is given by \cite{Kittel}
 \begin{equation}
 T_\mathrm{c} = \frac{2S(S + 1)}{3k_\mathrm{B}}zJ.
\label{eqn2}
\end{equation}
If the system is diluted by non-magnetic ions with concentration $n_\mathrm{p}$, $z$ can be written as $z = z_0(1-n_\mathrm{p})$, therefore $T_\mathrm{c0} = 2S(S+ 1)z_\mathrm{0}J/3k_\mathrm{B}$ (here $T_\mathrm{c0}$ and $z_0$ both refer to the undiluted system). In the present case, $T_\mathrm{c0} = 364.4$ K as is measured for our La$_{0.7}$Sr$_{0.3}$MnO$_3$ sample.

In hole-doped manganites, due to the mixed valence of Mn ions and the resulting coexistence of antiferromagnetic (AF) (Mn$^{3+}$-Mn$^{3+}$, Mn$^{4+}$-Mn$^{4+}$) SE and FM (Mn$^{3+}$-Mn$^{4+}$) DE, $T_\mathrm{c}$ should be better described as
\begin{equation}
T_\mathrm{c}=\frac{2S(S+1)}{3k_\mathrm{B}}\sum_{\alpha}z_{\alpha}J_{\alpha}.
\label{eqn3}
\end{equation}
Consequently, because of the selective nature of the dilution in our samples, the extrapolation of $T_\mathrm{c}(n_\mathrm{p})$ is supposed to intersect the $n_\mathrm{p}$-axis at a certain relative concentration $n_\mathrm{p} < 1$ \cite{Nam}. The tendency of $T_\mathrm{c}(n_\mathrm{p})$ to drop to zero at $n_\mathrm{p} = 1$ experimentally observed therefore unambiguously reflects the total dominance of FM DE interaction in these systems. The other AF SE interactions would be effectively negligible. The absence of AF SE interactions may be one reason behind the fact that La$_{0.7}$Sr$_{0.3}$MnO$_3$ has highest $T_\mathrm{c}$ among $AB$O$_3$ perovskite manganites \cite{Nam}. It is worth noting that $T_\mathrm{c}(n_\mathrm{p})$ would not follow the linear behavior up to $n_\mathrm{p} = 1$ as predicted by the MFT because there should exist a percolation threshold $n_\mathrm{c}$, above which clustering occurs and the ferromagnetic network collapses into superparamagnetic short-range ordering. Moreover, at high substitution concentrations, changes of crystalline structure and carrier concentration may become effective and dominate the magnetic properties.

\begin{figure}[t!]
\includegraphics[width=3.in]{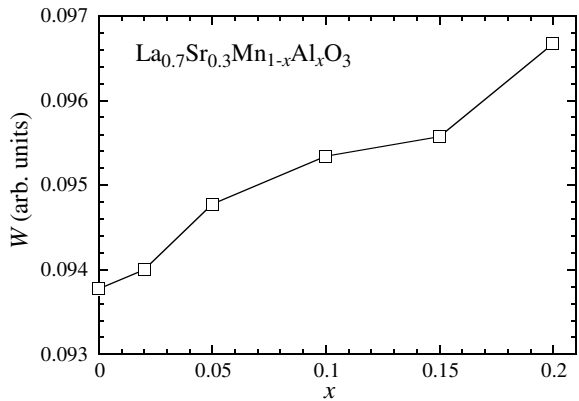}
\caption{Variation of the conduction electron bandwidth $W$ against Al concentration $x$. The monotonic increase of $W(x)$ is contradictory to the reduction in $T_\mathrm{c}(x)$ of La$_{0.7}$Sr$_{0.3}$Mn$_{1-x}$Al$_x$O$_3$.} \label{fig.4}
\end{figure}

In the case when both Al and Ti are substituted for Mn with according $n_\mathrm{p}$(Al) and $n_\mathrm{p}$(Ti), supposing that the FM DE is totally dominant in the system, the effective dilution concentration of the whole system is determined as
\begin{equation}
n_\mathrm{p} = 1-[1-n_\mathrm{p}(\mathrm{Al})][1-n_\mathrm{p}(\mathrm{Ti})].
\label{eqn4}
\end{equation}
The $T_\mathrm{c}(n_\mathrm{p})$ values of the $M^\prime= \mathrm{Al}_{0.07}\mathrm{Ti}_{0.03}$ ($n_\mathrm{p} = 0.19$) and $M^\prime= \mathrm{Al}_{0.14}\mathrm{Ti}_{0.06}$ ($n_\mathrm{p} = 0.36$) compounds are also added to Fig. \ref{fig.3}, showing a fairly well fit to the MFT prediction.
\section{CONCLUSIONS}
The magnetic properties of La$_{0.7}$Sr$_{0.3}$Mn$_{1-x}M^\prime_x$O$_3$, where Mn$^{3+}$ or Mn$^{4+}$ is selectively substituted by Al$^{3+}$ or Ti$^{4+}$, respectively, have been reexamined. The substitution appears to not only dilute the magnetic lattice, but also induce a redundancy of Mn ions. We have also discovered that, in terms of selective dilution and in the low substitution range, $T_\mathrm{c}$ linearly scales with $n_\mathrm{p}$, being in good agreement with the MFT approximation. The tendency of $T_\mathrm{c}$ to reduce to zero at $n_\mathrm{p} = 1$ suggests a totally dominant role of the DE mechanism in this system; the SE interaction is effectively negligible. The MFT analyzes are found to some extent also to be valid for the case when Al and Ti are both substituted for Mn. Our results also indicate that the variation of conduction electron bandwidth alone cannot explain the change of $T_\mathrm{c}$ in magnetically diluted manganites.
\begin{acknowledgments}
This work has been performed partly under the sponsorship of a collaborative project between the Institute of Materials Science (VAST, Vietnam) and Uppsala University (Sweden). A part of this work was carried out using facilities of the State Key Labs (IMS, VAST). Two of us would like to thank the University of Cincinnati and the National Science Foundation for support.
\end{acknowledgments}





\end{document}